\documentclass[
aps,%
12pt,%
final,%
notitlepage,%
oneside,%
onecolumn,%
nobibnotes,%
nofootinbib,%
superscriptaddress,%
showpacs,%
centertags]%
{revtex4}

\begin{document}

\title{Bases for qudits from a nonstandard approach to $SU(2)$}\footnote{From a talk presented at the 13th International Conference 
on Symmetry Methods in Physics (Dubna, Russia, 6-9 July 2009) organized in memory of Prof. Yurii Fedorovich Smirnov by the Bogoliubov 
Laboratory of Theoretical Physics of the JINR (Russia) and the ICAS at Yerevan State University (Armenia).}

\author{\firstname{Maurice R.}~\surname{Kibler}}


\affiliation{$^a$Universit\'e de Lyon, F--69622, Lyon, France \\ 
             $^b$Universit\'e Claude Bernard Lyon 1, Villeurbanne, France \\
             $^c$CNRS/IN2P3, Institut de Physique Nucl\'eaire de Lyon, France}


\begin{abstract}
Bases of finite-dimensional Hilbert spaces (in dimension $d$) of relevance for 
quantum information and quantum computation are constructed from angular momentum 
theory and $su(2)$ Lie algebraic methods. We report on a formula for deriving in one step 
the $(1+p)p$ qupits (i.e., qudits with $d = p$ a prime integer) of a complete set 
of $1+p$ mutually unbiased bases in ${\bf C}^p$. Repeated application of the formula 
can be used for generating mutually unbiased bases in ${\bf C}^d$ with $d = p^e$ 
($e \geq 2$) a power of a prime integer. A connection between mutually unbiased 
bases and the unitary group $SU(d)$ is briefly discussed in the case $d = p^e$.
\end{abstract}

\maketitle

\section{Prolegomena}

The use of symmetry adapted functions (or state vectors) is of paramount importance in 
molecular physics and condensed matter physics as well as in the clustering phenomenon 
of nuclei. For instance, wavefunctions adapted to a finite subgroup of $SU(2)$ turn out 
to be very useful in crystal- and ligand-field theory \cite{Smirnov2}-\cite{KiblerObninsk}. 

It is the purpose of the present work to report on state vectors adapted to the cyclic 
subgroup $C_d$ of $SO(3) \sim SU(2)/Z_2$. Such vectors give rise to bases of $SU(2)$, 
the so-called mutually unbiased bases (MUBs), to be defined in Section V. They play a 
fundamental role in quantum information and quantum computation in view of the fact 
that these bases describe qudits (the analogs of qubits in dimension $d$). A single formula 
for MUBs is obtained in this paper from a polar decomposition of $SU(2)$ and analysed 
in terms of quantum quadratic discrete Fourier transforms, Hadamard matrices, generalized 
quadratic Gauss sums, and special unitary groups $SU(d)$. 

Before dealing with the main body of this work, we continue this introduction with a brief 
survey for nonspecialists of some particular aspects of quantum computation and quantum 
information for which the concept of MUBs is useful. 

According to the law by Moore, the size of electronic and spintronic devices for a 
classical computer should approach 10 nm in 2018-2020, i.e., the scale where 
quantum effects are visible, a fact in favor of a quantum computer. This explains 
the growing interest for a new field, viz., the field of quantum information and 
quantum computation. Such a field, which started in the 1980's, is at the crossroads 
of quantum mechanics, discrete mathematics and informatics with the aim of building 
a quantum computer. We note in passing that, even in the case where the aim would 
not be reached, physics, mathematics, informatics and engineering will greatly benefit 
from the enormous amount of works along this line. 

In a quantum computer, classical 
bits (0 and 1) are replaced by quantum bits or qubits (that interpolate in some 
sense between 0 and 1). A qubit is a vector $| \phi \rangle$ in the two-dimensional 
Hilbert space ${\bf C}^2$:
\begin{eqnarray}
| \phi \rangle = x | 0 \rangle + y | 1 \rangle, 
\quad x \in {\bf C}, 
\quad y \in {\bf C}, 
\quad |x|^2 + |y|^2 = 1, 
\end{eqnarray}  
where $| 0 \rangle$ and $| 1 \rangle$ are the elements of an orthonormal basis 
in this space. The result of a measurement of $| \phi \rangle$ is not deterministic 
since it gives $| 0 \rangle$ or $| 1 \rangle$ with the probability $|x|^2$ or $|y|^2$, 
respectively. The consideration of $N$ qubits leads to work in the $2^N$-dimensional 
Hilbert space ${\bf C}^{2^N}$. Note that the notion of qubit, corresponding to 
${\bf C}^2$, is a particular case of the one of qudit, corresponding to ${\bf C}^d$ 
($d$ not necessarily in the form $2^N$). A system of $N$ qudits is associated with the 
Hilbert space ${\bf C}^{d^N}$. In this connection, the techniques developed for 
finite-dimensional Hilbert spaces are of paramount importance in quantum computation
and quantum calculation.

From a formal point of view, a quantum computer can be considered as a set of qubits, 
the state of which can be (controlled and) manipulated via unitary transformations. 
These transformations correspond to the product of elementary unitary operators called 
quantum gates acting on one or two qubits. Measurement of the qubits outcoming from a 
circuit of quantum gates yields the result of a (quantum) computation. In other words, 
a realization of quantum information processing can be performed by preparing a 
quantum system in a quantum state, then submitting this state to a unitary transformation 
and, finally, reading the outcome from a measurement.  

Unitary operator bases in ${\bf C}^{d}$ are of pivotal importance for quantum information 
and quantum computation as well as for quantum mechanics in general. The interest for unitary 
operator bases started with the seminal work by Schwinger \cite{Schwinger}. Among such bases, 
MUBs play a key role in quantum information and quantum computation 
\cite{Ivanovic}-\cite{Aschbacher}. From a very general point 
of view, MUBs are closely connected to the principle 
of complementarity introduced by Bohr in the early days of quantum mechanics. This 
principle, quite familiar in terms of observables like position and momentum, tells 
that for two noncommuting observables, if we have a complete knowledge of one observable, 
then we have a total uncertainty of the other. Equation (\ref{definition des mubs}) in Section V for 
$a \not= b$ indicates that the development in the basis $B_a$ of any vector of the basis 
$B_b$ is such that each vector of $B_a$ appears in the development with the probability 
$1/d$. This is especially interesting when translated in terms of measurements, the bases 
$B_a$ and $B_b$ corresponding to the (nondegenerate) eigenvectors of two noncommuting observables.

\section{A nonstandard basis for $SU(2)$}

The various irreducible representation classes of the group $SU(2)$ are 
characterized by a label $j$ with $2j \in {\bf N}$. The standard 
irreducible matrix representation associated with $j$ is spanned 
by the irreducible tensorial set 
   \begin{eqnarray}
B_{2j+1} = \{ |j , m \rangle : m = j, j-1, \ldots, -j  \},
   \end{eqnarray}
where the vector $|j , m \rangle$ is a common eigenvector of the Casimir 
operator $j^2$ and of the Cartan operator $j_z$ of the Lie algebra $su(2)$ 
of $SU(2)$. More precisely, we have the relations
          \begin{eqnarray}
          j^2 |j , m \rangle = j(j+1) |j , m \rangle, \quad 
          j_z |j , m \rangle = m      |j , m \rangle,
          \label{jdeux}
          \end{eqnarray}
which are familiar in angular momentum theory. (We use lower case letters for operators and capital letters 
for matrices so that $j^2$ in (\ref{jdeux}) stands for the square of a generalized angular momentum.)

Following the works in \cite{KiblerIJMPB, KiblerIJMPB2}, let 
us define the linear operators $v_{ra}$ and $h$ by
          \begin{eqnarray}
          v_{ra} = {e}^{{i} 2 \pi j r} |j , -j \rangle \langle j , j| 
                  + \sum_{m = -j}^{j-1} q^{(j-m)a} |j , m+1 \rangle \langle j , m| 
          \label{definition of vra} 
          \end{eqnarray}
and
   \begin{eqnarray}
   h = \sum_{m = -j}^j {\sqrt{ (j+m)(j-m+1) }} |j , m \rangle \langle j , m |, 
   \label{definition of h}
   \end{eqnarray}
where  
          \begin{eqnarray}       
          r \in {\bf R}, \quad
          q = e^{2 \pi {i} / (2j+1)}, \quad 
          a \in {\bf R}.
          \label{parameters} 
          \end{eqnarray}
It can be checked that the three operators
	        \begin{eqnarray}
  j_+ = h           v_{ra},    \quad  
  j_- = (v_{ra})^{\dagger} h,  \quad 
  j_z = \frac{1}{2} \left[ h^2 - (v_{ra})^{\dagger} h^2 v_{ra} \right], 
          \label{polar decomposition}
          \end{eqnarray}
where $(v_{ra})^{\dagger}$ stands for the adjoint of $v_{ra}$, satisfy the commutation relations 
     \begin{eqnarray}
  \left[ j_z,j_{+} \right] = + j_{+},  \quad 
  \left[ j_z,j_{-} \right] = - j_{-},  \quad 
  \left[ j_+,j_- \right]   = 2j_z 
     \label{adL su2}
     \end{eqnarray}
of the algebra $su(2)$. (In angular momentum theory, the operators $j_{+}$ and 
$j_{-}$ are connected to $j^2$ via $j^2 = j_{\pm} j_{\mp} + j_z(j_z \mp 1)$.)

The operator $v_{ra}$ is unitary while the operator $h$ is Hermitian. Thus, 
Eq. (\ref{polar decomposition}) corresponds to a polar decomposition of $su(2)$ 
with the help of the operators $v_{ra}$ and $h$. It should be noted that  
$v_{ra}$ can be derived in terms of operators acting on the tensor product 
of two commuting quon algebras associated with two truncated harmonic 
oscillators. The latter oscillators play a central role in the intoduction 
of $k$-fermions which are supersymmetric objects interpolating between 
fermions and bosons \cite{DaoHasKib, DaoHasKib2}.

It is evident that $v_{ra}$ and $j^2$ commute. Therefore, the $\{ j^2 , v_{ra} \}$ scheme 
constitutes an alternative to the $\{ j^2 , j_z \}$ scheme. This yields the following result. 

\bigskip

       {\bf Result 1.} {\it For fixed $j$, $r$ and $a$, the $2j+1$ vectors 
          \begin{eqnarray}
|j \alpha ; r a \rangle = \frac{1}{\sqrt{2j+1}} \sum_{m = -j}^{j} 
q^{(j + m)(j - m + 1)a / 2 - j m r + (j + m)\alpha} | j , m \rangle, 
          \label{j alpha r a in terms of jm}
          \end{eqnarray} 
with $\alpha = 0, 1, \ldots, 2j$, are common eigenvectors of $v_{ra}$ and $j^2$. The 
eigenvalues of $v_{ra}$ are given by 
      \begin{eqnarray}
v_{ra} |j \alpha ; r a \rangle = q^{j(a+r) - \alpha} |j \alpha ; r a \rangle,  
      \label{evp de vra}
      \end{eqnarray}
so that the spectrum of $v_{ra}$ is nondegenerate.}

\bigskip

For fixed $j$, $r$ and $a$, the inner product 
      \begin{eqnarray}
\langle j \alpha ; r a | j \beta ; r a \rangle = \delta_{\alpha,\beta}
      \label{jalphabetara}
      \end{eqnarray}
shows that $\{ |j \alpha ; r a \rangle : \alpha = 0, 1, \ldots, 2j \}$ 
is an orthonormal set which provides a nonstandard basis for the 
irreducible matrix representation of $SU(2)$ associated with $j$. 

\section{Quantum quadratic discrete Fourier transform}

In view of the interest of the bases $\{ |j \alpha ; 0 a \rangle : \alpha = 0, 1, \ldots, 2j \}$ 
for quantum information and quantum computation, we shall continue with the case $r = 0$. From 
now on, we shall also assume that $a = 0, 1, \ldots, 2j$. Furthermore, by making the following 
change of notation 
      \begin{eqnarray}
n \equiv j+m, \quad | n \rangle \equiv |j , m \rangle, \quad d \equiv 2j+1,  
      \label{change1}
      \end{eqnarray}
Eq.~(\ref{j alpha r a in terms of jm}) gives 
          \begin{eqnarray}
|j \alpha ; 0 a \rangle = \frac{1}{\sqrt{d}} \sum_{n=0}^{d-1} 
q^{n(d-n)a / 2 + n \alpha} | n \rangle. 
          \label{j alpha in terms of n}
          \end{eqnarray} 
Alternatively, the change of notation
      \begin{eqnarray}
k \equiv j-m, \quad | k \rangle \equiv |j , m \rangle, \quad d \equiv 2j+1  
      \label{change2}
      \end{eqnarray}
leads to 
          \begin{eqnarray}
|j \alpha ; 0 a \rangle = \frac{1}{\sqrt{d}} \sum_{k=0}^{d-1} 
q^{(k+1)(d-k-1)a / 2 - (k+1) \alpha} | k \rangle. 
          \label{j alpha in terms of k}
          \end{eqnarray}
Equations (\ref{j alpha in terms of n}) and (\ref{j alpha in terms of k}) were used in 
\cite{AlbouyKibler} and \cite{Kibler0809, K0809deux}. They are equivalent as far as quadratic 
discrete Fourier transforms and mutually unbiased bases (MUBs) are concerned. Both 
Eqs.~(\ref{j alpha in terms of n}) and (\ref{j alpha in terms of k}) correspond to 
quantum quadratic discrete Fourier transforms which can be inverted to give 
          \begin{eqnarray}
| n \rangle = \frac{1}{\sqrt{d}} q^{-n(d-n)a / 2} \sum_{\alpha=0}^{d-1} 
q^{-\alpha n} |j \alpha ; 0 a \rangle  
          \end{eqnarray} 
and
          \begin{eqnarray}
| k \rangle = \frac{1}{\sqrt{d}} q^{-(k+1)(d-k-1)a / 2} \sum_{\alpha=0}^{d-1} 
q^{\alpha(k+1)} |j \alpha ; 0 a \rangle.
          \end{eqnarray}
Note that the word {\it quantum} in {\it quantum quadratic discrete Fourier transform} refers to the fact 
that the vectors $| n \rangle$ or $| k \rangle$ (corresponding to $|j , m \rangle$) are used in the quantum 
theory of generalized angular momentum. 

In the following we shall adopt the change of notation (\ref{change1}) and shall re-define $|j \alpha ; 0 a \rangle$ 
as $| a \alpha \rangle$. In other words 
          \begin{eqnarray}
| a \alpha \rangle = \frac{1}{\sqrt{d}} \sum_{n=0}^{d-1} q^{n(d-n)a / 2 + n \alpha} | n \rangle 
          \label{aalpha}
          \end{eqnarray}
or, in an equivalent way, 
          \begin{eqnarray}
| n \rangle        = \frac{1}{\sqrt{d}} q^{-n(d-n)a / 2} \sum_{\alpha=0}^{d-1} q^{-\alpha n} | a \alpha \rangle. 
          \end{eqnarray}   
Note that the action of the operator $v_{0a}$ on the vector $| n \rangle$ reads
          \begin{eqnarray}
v_{0a} |n \rangle = q^{-(n+1)a} |n+1 \rangle
          \end{eqnarray}
modulo $d$. It is clear that the basis 
          \begin{eqnarray}
          B_{0a} = \{ | a \alpha \rangle : \alpha = 0, 1, \ldots, d-1 \}
          \label{baseB0a}
          \end{eqnarray}
is an alternative to the basis $B_d \equiv B_{2j+1}$. There 
are $d=2j+1$ bases of this type for $a$ in the ring ${\bf Z}/d{\bf Z}$. Each 
basis $B_{0a}$ spans the regular representation of the cyclic group $C_d$. 

\section{Quadratic discrete Fourier transform}
The expression
          \begin{eqnarray}
\left( H_{0a} \right)_{n \alpha} = \frac{1}{\sqrt{d}} q^{n(d-n)a / 2 + n \alpha}, 
          \label{element of Ha}
          \end{eqnarray} 
occurring in (\ref{aalpha}), defines the ${n \alpha}^{\rm th}$ matrix element of a quadratic discrete 
Fourier transform. To be more precise, for fixed $d$ and $a$, let us consider the transformation  
    \begin{eqnarray}	                  
x = \{ x(n     ) \in {\bf C} : n      = 0, 1, \ldots, d-1 \} \leftrightarrow
y = \{ y(\alpha) \in {\bf C} : \alpha = 0, 1, \ldots, d-1 \} 
	  \end{eqnarray} 
defined by 
    \begin{eqnarray}	                  
y(\alpha) = \sum_{n = 0}^{d-1} \left( H_{0a} \right)_{n \alpha} x(n) \Leftrightarrow
x(n) = \sum_{\alpha = 0}^{d-1} \overline{\left( H_{0a} \right)_{n \alpha}} y(\alpha). 
    \label{transfo yx et xy en Ha} 
	  \end{eqnarray} 
The particular case $a=0$ corresponds to the ordinary discrete Fourier transform which satisfies
    \begin{eqnarray}	
    \left( H_{00} \right)^4 = I_d,                 
	  \end{eqnarray} 
where $I_d$ is the identity $d \times d$ matrix. For $a \not= 0$, the bijective transformation 
$x \leftrightarrow y$ can be thought of as a quadratic discrete Fourier transform. The analog 
of the Parseval-Plancherel theorem for the usual Fourier transform can be expressed in the 
following way. 

\bigskip

       {\bf Result 2.} {\it The quadratic discrete Fourier transforms 
$x \leftrightarrow y$ and $x' \leftrightarrow y'$ associated with the same 
matrix $H_{0a}$, $a \in {\bf Z}/d{\bf Z}$, satisfy the conservation rule 
    \begin{eqnarray}	                  
\sum_{\alpha = 0}^{d-1} \overline{y(\alpha)} y'(\alpha) = \sum_{n = 0}^{d-1} \overline{x(n)} x'(n),
    \label{Parseval-Plancherel} 
	  \end{eqnarray}
where the common value is independent of $a$.}

\bigskip

It is to be observed that the matrix $H_{0a}$ is a generalized Hadamard matrix in the sense that the modulus 
of each of its matrix element is equal to $1/\sqrt{d}$. Such a matrix reduces the endomorphism associated 
with the operator $v_{0a}$. As a matter of fact, we have
      \begin{eqnarray}
\left( H_{0a} \right)^{\dagger} V_{0a} H_{0a} = 
q^{(d-1)a / 2} \pmatrix{
q^{1}                &      0 &       \ldots &          0    \cr
0                    & q^{2}  &       \ldots &          0    \cr
\vdots               & \vdots &       \ldots &     \vdots    \cr
0                    &      0 &       \ldots & q^{d}         \cr
},
     \label{endomor}
     \end{eqnarray}
where the matrix
      \begin{eqnarray}
V_{0a} = 
\pmatrix{
0                    &    q^a &      0  & \ldots &          0    \cr
0                    &      0 & q^{2a}  & \ldots &          0    \cr
\vdots               & \vdots & \vdots  & \ldots &     \vdots    \cr
0                    &      0 &      0  & \ldots & q^{(d-1)a}    \cr
1                    &      0 &      0  & \ldots &          0    \cr
}
     \label{matrix V0a}
     \end{eqnarray}
represents the linear operator $v_{0a}$ on the basis 
      \begin{eqnarray}
B_d = \{ | n \rangle : n = d-1, d-2, \ldots, 0 \},     
      \end{eqnarray}
known as the computational basis in quantum information and quantum computation. 

The Hadamard matrices $H_{0a}$ and $H_{0b}$ ($a, b \in {\bf Z}/d{\bf Z}$) are connected 
to the inner product $\langle a \alpha | b \beta \rangle$. In fact, we have
      \begin{eqnarray}
\left( \left( H_{0a} \right)^{\dagger} H_{0b} \right)_{\alpha \beta} 
= \langle a \alpha | b \beta \rangle 
= \frac{1}{d} \sum_{n = 0}^{d-1} q^{n(d-n)(b-a) / 2 + n(\beta - \alpha)}. 
     \end{eqnarray}
Thus, each matrix element of $\left( H_{0a} \right)^{\dagger} H_{0b}$ can be written 
in the form of a generalized quadratic Gauss sum $S(u, v, w)$ defined by \cite{BerndtEW} 
     \begin{eqnarray}
     S(u, v, w) = \sum_{n = 0}^{|w|-1} e^{i \pi (u n^2 + v n) / w},
     \label{Gauss}
     \end{eqnarray}
where $u$, $v$, and $w$ are integers such that $u$ and $w$ are mutually prime, 
$uw \not= 0$, and $uw + v$ is even. In detail, we obtain
     \begin{eqnarray}
\langle a \alpha | b \beta \rangle 
= \left( \left( H_{0a} \right)^{\dagger} H_{0b} \right)_{\alpha \beta} 
= \frac{1}{d} S(u, v, w),   
     \label{ee}
     \end{eqnarray}
with the parameters
     \begin{eqnarray}
     u = a - b, \quad v = -(a - b)d - 2(\alpha - \beta), \quad w = d,
     \label{ff}
     \end{eqnarray}
which ensure that $uw + v$ is even. 

The matrix $V_{0a}$ can be decomposed as 
     \begin{eqnarray}
V_{0a} = X_0 Z^a,
     \end{eqnarray}
where
       \begin{eqnarray}
X_0 = 
\pmatrix{
0                    &      1 &      0  & \ldots &       0 \cr
0                    &      0 &      1  & \ldots &       0 \cr
\vdots               & \vdots & \vdots  & \ldots &  \vdots \cr
0                    &      0 &      0  & \ldots &       1 \cr
1                    &      0 &      0  & \ldots &       0 \cr
}
        \end{eqnarray}
and
        \begin{eqnarray}
Z = 
\pmatrix{
1                    &      0 &      0    & \ldots &       0       \cr
0                    &      q &      0    & \ldots &       0       \cr
0                    &      0 &      q^2  & \ldots &       0       \cr
\vdots               & \vdots & \vdots    & \ldots &  \vdots       \cr
0                    &      0 &      0    & \ldots &       q^{d-1} \cr
}.
        \end{eqnarray}
The unitary matrices $X_0$ and $Z$ $q$-commute in the sense that 
     \begin{eqnarray}
X_0 Z - q Z X_0 = 0. 
     \label{qcom of Xr and Z}
     \end{eqnarray}
In addition, they satisfy
     \begin{eqnarray}
(X_0)^d = Z^d = I_d.  
     \label{dpower of Xr and Z}
     \end{eqnarray}
Equations (\ref{qcom of Xr and Z}) and 
(\ref{dpower of Xr and Z}) show that $X_0$ (to be noted as $X$ in what follows in order to conform 
to the notations used for Pauli matrices) and $Z$ constitute a Weyl pair. Weyl pairs were introduced 
at the beginning of quantum mechanics \cite{Weyl} and used for building operator unitary bases 
\cite{Schwinger}. The Weyl pair ($X , Z$) turns out to be an integrity basis for generating a set 
$\{ X^a Z^b : a,b \in {\bf Z}/d{\bf Z} \}$ of $d^2$ generalized Pauli matrices in $d$ dimensions 
(see for instance \cite{Kibler0809, K0809deux} and references therein). In this 
respect, note that for $d=2$ we have 
   \begin{eqnarray}
X      =     \sigma_x, \quad   
Z      =     \sigma_z, \quad 
XZ     = - i \sigma_y, \quad 
X^0Z^0 =     \sigma_0,
   \end{eqnarray}
in terms of the ordinary Pauli matrices $\sigma_0 = I_2$, $\sigma_x$, $\sigma_y$, and $\sigma_z$. Equations 
(\ref{qcom of Xr and Z}) and (\ref{dpower of Xr and Z}) can be generalized through 
     \begin{eqnarray}
V_{0a} Z - q Z V_{0a} = 0, \quad (V_{0a})^d = e^{i \pi (d-1)a} I_d, \quad Z^d = I_d, 
     \end{eqnarray}
so that other pairs of Weyl can be obtained from $V_{0a}$ and $Z$. 

\section{Mutually unbiased bases}
From a very general point of view, let us recall that two orthonormal 
bases $B_a = \{ | a \alpha \rangle : \alpha = 0, 1, \ldots, d-1 \}$ and 
      $B_b = \{ | b \beta  \rangle : \beta  = 0, 1, \ldots, d-1 \}$ of the Hilbert space 
${\bf C}^{d}$ are said to be mutually unbiased if and only if the inner product 
$\langle a \alpha | b \beta \rangle$ has a modulus independent of $\alpha$ and 
$\beta$. In other words
          \begin{eqnarray}
\forall \alpha \in {\bf Z}/d{\bf Z}, 
\forall \beta  \in {\bf Z}/d{\bf Z} : 
| \langle a \alpha | b \beta \rangle | = \delta_{a , b}
\delta_{\alpha , \beta} + (1 - \delta_{a , b}) \frac{1}{\sqrt{d}}.
          \label{definition des mubs}
          \end{eqnarray}          
From Eq.~(\ref{definition des mubs}), note that if two MUBs undergo the same unitary or 
antiunitary transformation, they remain mutually unbiased. It is well-known that the 
maximum number ${\cal N}$ of MUBs in ${\bf C}^d$ is ${\cal N} = 1+d$ and that this number is 
attained when $d$ is a prime number $p$ or a power $p^e$ ($e \geq 2$) of a prime number $p$ 
\cite{Ivanovic}-\cite{BandyoGPM5}. In the other cases 
($d \not= p^e$, $p$ prime and $e$ integer with $e \geq 1$), the number ${\cal N}$ 
is not known although it can be shown that $3 \leq {\cal N} \leq 1+d$  (see for 
example \cite{Grassl}). In the general composite case $d = \prod_i p_i^{e_i}$, 
we know that $1 + {\rm min}(p_i^{e_i}) \leq {\cal N} \leq 1+d$ (see for example 
\cite{Aschbacher}).

\subsection{Mutually unbiased bases for $d$ prime}

For $d=2$, it can be checked that the bases $B_{00}$, $B_{01}$, and $B_2$ 
are $1+d = 3$ MUBs. A similar result follows for $d=3$: the bases $B_{00}$, 
$B_{01}$, $B_{02}$, and $B_3$ are $1 + d = 4$ MUBs. This can be generalized 
by the following result. 

\bigskip

       {\bf Result 3.} {\it For $d=p$, with $p$ a prime number, the bases $B_{00}, B_{01}, \ldots, B_{0p-1}, B_{p}$ 
form a complete set of $1+p$ MUBs. The $p^2$ vectors $| a \alpha \rangle$, with 
$a, \alpha = 0, 1, \ldots, p-1$, of the bases  $B_{00}, B_{01}, \ldots, B_{0p-1}$ are given by a single formula 
(namely Eq.~(\ref{aalpha})).}

\bigskip

The proof of Result 3 is as follows. First, Eq.~(\ref{aalpha}) yields  
    \begin{eqnarray}	                  
 | \langle k | a \alpha \rangle | = \frac{1}{\sqrt{p}}, 
	  \end{eqnarray} 
a relation that holds for all $k$, $a$, and $\alpha$ in ${\bf Z}/p{\bf Z}$ so that each basis $B_{0a}$ is unbiased 
with $B_{p}$. Second, the generalized quadratic Gauss sum $S(u,v,w)$ in (\ref{ee}), with $d=p$ prime, can be calculated to give
     \begin{eqnarray}
 | \langle a \alpha | b \beta \rangle | = \frac{1}{\sqrt{p}},  
     \end{eqnarray}
for all $a$, $b$, $\alpha$, and $\beta$ in ${\bf Z}/p{\bf Z}$. This completes the proof. 

We note in passing that, in the case where $d = p$ is a prime integer, the 
product $\left( H_{0a} \right)^{\dagger} H_{0b}$ is another generalized Hadamard matrix.

To close this subsection, we may ask what becomes Result 3 when the prime integer $p$ is replaced by an  
arbitrary (not prime) integer $d$. In this case, the formula (\ref{aalpha}) does not provide a complete 
set of $1+d$ MUBs. However, it is possible to show \cite{Kibler0809, K0809deux} that the bases $B_{0a}$, $B_{0 a \oplus 1}$, 
and $B_d$ are three MUBs in ${\bf C}^d$ (the addition $\oplus$ is understood modulo 
$d$). This result is in agreement with the well-known result according to which the maximum number 
of MUBs in ${\bf C}^d$, with $d$ arbitrary, is greater or equal to 3 (\cite{Grassl}). Moreover, 
it can be proved \cite{Kibler0809, K0809deux} that the bases $B_{0a}$ and $B_{0 a \oplus 2}$ are unbiased for $d$ odd with 
$d \geq 3$ ($d$ prime or nor prime).

\subsection{Mutually unbiased bases for $d$ power of a prime}

Equation (\ref{aalpha}) can be used for deriving a complete set of $1+p^e$ MUBs in the case where 
$d=p^e$ is a power ($e \geq 2$) of a prime integer $p$. The general case is very much involved. Hence, 
we shall proceed with the example $p=e=2$ corresponding to two qubits. 

For $d = 2^2 = 4$, the application of (\ref{aalpha}) and (\ref{baseB0a}) yields four bases $B_{0a}$ 
($a = 0, 1, 2, 3$). As a point of fact, the bases $B_{00}$, $B_{01}$, $B_{02}$, $B_{03}$, and $B_4$ 
do not form a complete set of $1+d = 5$ MUBs. However, it is possible to construct a set of five MUBs 
from repeated application of (\ref{aalpha}).

Four of the five MUBs for 
$d=4$ can be constructed from the direct products 
$|a \alpha \rangle \otimes |b \beta \rangle$ which 
are eigenvectors of the operators 
$v_{0a} \otimes v_{0b}$. Obviously, the set 
   	\begin{eqnarray}
B_{0a0b} = \{ |a \alpha \rangle \otimes |b \beta \rangle : \alpha, \beta = 0, 1 \}
   	\end{eqnarray} 
is an orthonormal basis in ${\bf C}^4$. It is evident that $B_{0000}$ and $B_{0101}$ are two unbiased bases 
since the modulus of the inner product of $|0 \alpha \rangle \otimes |0 \beta \rangle$ by $|1 \alpha' \rangle \otimes |1 \beta' \rangle$
is 
   	\begin{eqnarray}
| \langle 0 \alpha | 1 \alpha' \rangle \langle 0 \beta | 1 \beta' \rangle | = \frac{1}{\sqrt{4}}.
   	\end{eqnarray}    	
A similar result holds for the two bases $B_{0001}$ and $B_{0100}$. However, 
the four bases $B_{0000}$, $B_{0101}$, $B_{0001}$, and $B_{0100}$ are not 
mutually unbiased. A possible way to overcome this no-go result is to keep 
the bases $B_{0000}$ and $B_{0101}$ intact and to re-organize the vectors 
inside the bases $B_{0001}$ and $B_{0100}$ in order to obtain four MUBs. We 
are thus left with four bases 
   	\begin{eqnarray}
W_{00} \equiv B_{0000}, \quad W_{11} \equiv B_{0101}, \quad W_{01}, \quad W_{10}, 
   	\end{eqnarray}   
which together with the computational basis $B_4$ give five MUBs. In detail, we have 
   	\begin{eqnarray}
W_{00} & = & \{         |0 \alpha \rangle \otimes |0 \beta \rangle : \alpha, \beta = 0, 1 \},  \\
W_{11} & = & \{         |1 \alpha \rangle \otimes |1 \beta \rangle : \alpha, \beta = 0, 1 \},  \\
W_{01} & = & \{ \lambda |0 \alpha \rangle \otimes |1 \beta \rangle + 
                     \mu |0 \alpha \oplus 1 \rangle \otimes |1 \beta \oplus 1 \rangle : \alpha, \beta = 0, 1 \},  \\
W_{10} & = & \{ \lambda |1 \alpha \rangle \otimes |0 \beta \rangle + 
                     \mu |1 \alpha \oplus 1 \rangle \otimes |0 \beta \oplus 1 \rangle : \alpha, \beta = 0, 1 \},  
    \label{les quatre W}
   	\end{eqnarray} 
where
   	\begin{eqnarray}
\lambda = \frac{1-i}{2}, \quad \mu = \frac{1+i}{2}
   	\end{eqnarray} 
and the vectors of type $|a \alpha \rangle$ are given by the master formula (\ref{aalpha}). As a r\'esum\'e, only 
two formulas are necessary for obtaining the $d^2 = 16$ vectors $| a b ; \alpha \beta \rangle$ for the bases $W_{ab}$, namely 
   	  \begin{eqnarray}
W_{00}, W_{11} & : & | a a ; \alpha \beta \rangle = |a \alpha \rangle \otimes |a \beta \rangle, 
\label{W00W11} \\
W_{01}, W_{10} & : & | a a \oplus 1 ; \alpha \beta \rangle = \lambda |a \alpha          \rangle \otimes |a \oplus 1 \beta \rangle + 
                                                                 \mu |a \alpha \oplus 1 \rangle \otimes |a \oplus 1 \beta \oplus 1 \rangle, 
      \label{W01W10}
   	  \end{eqnarray} 
for all $a, \alpha$, and $\beta$ in ${\bf Z}/2{\bf Z}$.

It is to be noted that the vectors of the $W_{00}$ and $W_{11}$ bases are not intricated 
(i.e., each vector is the direct product of two vectors) while the vectors of the $W_{01}$ 
and $W_{10}$ bases are intricated (i.e., each vector is not the direct product of two 
vectors).

Generalization of (\ref{W00W11}) and (\ref{W01W10}) can be obtained in more complicated situations 
(two qupits, three qubits, \ldots). The generalization of (\ref{W00W11}) is immediate. The generalization 
of (\ref{W01W10}) can be achieved by taking linear combinations of vectors such that each linear combination 
is made of vectors corresponding to the same eigenvalue of the relevant tensor product of operators of type $v_{0a}$.

\section{Mutually unbiased bases and unitary groups}

In the case where $d$ is a prime integer or a power of a prime integer, 
it is known that the set $\{ X^aZ^b : a, b = 0, 1, \ldots, d-1 \}$ of cardinality $d^2$ can be partitioned into 
$1+d$ subsets containing each $d-1$ commuting matrices (cf. \cite{BandyoGPM5}). By way of illustration, 
for $d=p$ with $p$ prime, the $1+p$ sets of $p-1$ commuting matrices 
are easily seen to be                 
           \begin{eqnarray}  
{\cal V}_0       &=      &  \{ X^0 Z^a        :  a = 1, 2, \ldots, p-1 \}, \\
{\cal V}_1       &=      &  \{ X^a Z^0        :  a = 1, 2, \ldots, p-1 \}, \\ 
{\cal V}_2       &=      &  \{ X^a Z^a        :  a = 1, 2, \ldots, p-1 \}, \\
{\cal V}_3       &=      &  \{ X^a Z^{2a}     :  a = 1, 2, \ldots, p-1 \}, \\
                 &\vdots  &                                                \nonumber \\
 {\cal V}_{p-1}  &=      &  \{ X^a Z^{(p-2)a} :  a = 1, 2, \ldots, p-1 \}, \\
 {\cal V}_{p}    &=      &  \{ X^a Z^{(p-1)a} :  a = 1, 2, \ldots, p-1 \}. 
           \end{eqnarray} 
Each of the $1+p$ sets ${\cal V}_0, {\cal V}_1, \ldots, {\cal V}_{p}$ can be put in a one-to-one correspondance 
with one basis of the complete set of $1+p$ MUBs. In fact, ${\cal V}_0$ is associated with the computational basis 
while ${\cal V}_1, {\cal V}_2, \ldots, {\cal V}_{p}$ are associated with the $p$ remaining MUBs in view of 
           \begin{eqnarray}  
V_{0 a} \in {\cal V}_{a \oplus 1}, \quad a = 0, 1, \ldots, p-1. 
           \end{eqnarray} 
Keeping into account the fact that the set $\{ X^a Z^b : a,b = 0, 1, \ldots, p-1 \} \setminus \{ X^0 Z^0 \}$
spans the Lie algebra of $SU(p)$, we get the following result.

\bigskip

       {\bf Result 4.} {\it For $d=p$, with $p$ a prime integer, the Lie algebra 
$su(p)$ of the group $SU(p)$ can be decomposed into a direct sum of $1+p$ abelian 
subalgebras each of dimension $p-1$, i.e.
                  \begin{eqnarray}
{su}(p) \simeq 
{ v}_0     \uplus 
{ v}_1     \uplus 
\ldots     \uplus      
{ v}_{p}     
                  \end{eqnarray}
where the $1+p$ subalgebras ${ v}_0, { v}_1, \ldots, { v}_p$ are 
Cartan subalgebras generated respectively by the sets ${\cal V}_0, {\cal V}_1, \ldots, {\cal V}_{p}$ 
containing each $p - 1$ commuting matrices.}

\bigskip

Result 4 can be extended when $d = p^e$ with $p$ a prime integer and 
$e$ an integer ($e \geq 2$): there exists a decomposition of $su(p^e)$ into $1+p^e$ 
abelian subalgebras of dimension $p^e - 1$ \cite{KKU}-\cite{autresdecomp2} 
(see also \cite{Kibler0809, K0809deux}). 

\section{Closing remarks}

MUBs prove to be useful in classical information theory (network communication protocols), 
in quantum information theory (quantum state tomography and quantum cryptography), and in 
the theory of quantum mechanics as for the solution of the Mean King problem  
and the understanding of the Feynman path integral formalism (see \cite{Kibler0809, K0809deux} 
for an extensive list of references). 

There exist numerous ways of constructing sets of MUBs. Most of them are based 
on discrete Fourier transform over Galois fields and Galois rings, discrete Wigner 
distribution, generalized Pauli operators, mutually orthogonal Latin squares, discrete 
geometry methods, angular momentum theory and Lie-like approaches. In many of the papers 
dealing with the construction of MUBs for $d$ a prime integer or a power 
of a prime integer, the explicit derivation of the bases requires the diagonalisation of a set 
of matrices. 

In the present paper, the generic formula (\ref{aalpha}) arises from the 
diagonalisation of a single matrix (the matrix $V_{0a}$), for the ${\cal N} = 1+p$ MUBs corresponding to $d= p$ with $p$ 
a prime integer. Repeated application ($e$ times) of this formula can be used in the case where 
$d = p^e$ is the power of a prime integer. Results 1 and 3 of this paper concern the closed form formula (\ref{aalpha}).  
Its derivation is based on the master matrix 
$V_{0a}$. From $V_{0a}$ we can deduce the Weyl pair ($X , Z$) through
    \begin{eqnarray}
X = V_{00}, \quad Z = \left( V_{00} \right)^{\dagger} V_{01}.
   	\end{eqnarray}
The operators $X$ and $Z$ are known as the flip or shift and clock operators, 
respectively. For $d$ arbitrary, they are at the root of the Pauli 
group, a finite subgroup of order $d^3$ of the group $SU(d)$, of 
considerable importance in quantum information and quantum 
computation (e.g., see \cite{geometrical1}-\cite{Albouythesis} 
and references therein for recent geometrical approaches 
to the Pauli group). The Pauli group is relevant for describing 
quantum errors and quantum fault tolerance in quantum computation.
 
\section*{Acknowledgments}
This paper is dedicated to the memory of Yurii Fedorovich Smirnov. It was presented at the 13th 
International Conference on Symmetry Methods in Physics (SYMPHYS-XIII) organized in memory of 
Prof.~Yurii Fedorovich Smirnov by the Bogoliubov Laboratory of Theoretical Physics of the Joint 
Institute for Nuclear Research and the International Center for Advanced Studies at Yerevan State 
University (the conference was held in Dubna, Russia, 6-9 July 2009). The author acknowledges the 
Organizing Committee of SYMPHYS-XIII, especially George S. Pogosyan, for their kind invitation to 
participate to this interesting conference.

\end{document}